\documentclass[]{aa}
\usepackage{epsf,psfig,epsfig,graphics}
\begin{document}
    \title{GMRT observations of the field of INTEGRAL X-ray sources- II\\
              {\it (newly discovered hard X-ray sources)}}
             
    \author{M. Pandey\inst{1}, A. P. Rao\inst{2}, R. Manchanda\inst{3}, P. Durouchoux\inst{4}, 
C. H. Ishwara-Chandra.\inst{2}}

    \offprints{M. Pandey : mamta@ncra.tifr.res.in}

\institute{Department of Physics, Mumbai University, Mumbai - 400 098\and
NCRA, TIFR, Post Bag 3, Ganeshkhind, Pune - 411 007\and
Department of Astronomy and Astrophysics, TIFR, Colaba, Mumbai -400005\and
CNRS FRE 2591/CEA Saclay, DSM/DAPNIA/SAP, F-91191 Gif sur Yvette Cedex, France}
 \date{Received ; accepted}

\authorrunning{M. Pandey et al.}
\titlerunning{GMRT Observations of the field of INTEGRAL X-ray sources- II}

\abstract{We have conducted low-frequency radio observations with the Giant 
Metrewave Radio Telescope (GMRT) of 40 new hard X-ray sources discovered 
by the INTEGRAL satellite. This survey was conducted in order, to study radio 
emissions from these sources, to provide precise position and to identify new microquasar 
candidates. From our observations 
we find that 24 of the X-ray sources have radio candidates within the INTEGRAL error circle. 
Based on the radio morphology, variability and information available from different wavelengths, 
we categorize them as seventeen Galactic sources (4 unresolved, 7 extended, 6 extended sources in 
diffuse region) and seven extragalactic sources (2 unresolved, 5 extended). Detailed account for 
seventeen of these sources was presented in earlier paper. Based on the radio data for the remaining sources 
at 0.61 GHz, and the available information from NVSS, DSS, 
2MASS and NED, we have identified possible radio counterparts for the hard X-ray sources. 
The three unresolved sources, viz IGR J17303$-$0601, IGR J17464$-$3213, and IGR J18406$-$0539 
are discussed in detail. These sources have been identified as X-ray binaries with compact central 
engine and variable in X-ray and in the radio, and are most likely microquasar 
candidates. The remaining fourteen sources have extended radio morphology and are either diffuse 
Galactic regions or extragalactic in origin.        
 
-- Techniques : interferometry}

    \maketitle

\section{Introduction}
Many new hard X-ray emitting sources have been discovered during the deep
Galactic plane survey by {\emph{INTEGRAL}} satellite mission. The IBIS instrument
has a point source location accuracy (PSLA) of typically 1 -- 3$'$ within a large
field of view 29$^\circ$ $\times$ 29$^\circ$ (Ubertini et al. 2003). 55 new
hard X-ray sources have been reported in literature. A majority of these sources are
believed to be Galactic X-ray binaries  with a compact object
orbiting a companion star (Bird et al. 2004). Some of these sources are identified as AGNs, radio galaxies,
pulsars, CVs and dwarf nova. A detailed study of X-ray sources in the multi-wavelength band is essential to 
understand the emission mechanism and the  accretion process on to the compact companion  neutron stars
(NS) or black holes (BH). The radio imaging of these sources can establish whether some of these are
 radio  emitting X-ray binaries (REXBs)  and show any  microquasar like features. Due to their similarity
with quasars, the jet feature in microquasars provide important information about the underlying
physical phenomenon and the possible disk-jet connection which may power the observed emission in different
wave bands. Their X-ray, infrared and radio properties can lead to classification schemes.

\par
The detailed radio observations for these sources were made immediately after their discovery using the GMRT,
to find the possible radio counterparts within the location error box of
the X-ray source, to measure precise position if detected and to study the low
frequency radio nature of the hard X-ray sources. The radio morphology of a source also
provides its identification, viz Galactic, which are mainly compact (Becker et al. 1990) or extragalactic 
mostly extended (Jackson 1999).  At meter wavelengths REXBs show point source morphology (Pandey {et al.} 2005a).
We have also analyzed the available NVSS images at 1.4 GHz and
 other  radio data in order to understand
the radio spectrum of these sources. The archival data, from DSS, 2MASS and NED images is also used in our analysis 
to facilitate a complete study of these sources.
\begin{table*}[ht]
\begin{center}
\caption{List of target INTEGRAL sources observed with GMRT}
\begin{tabular}{p{25.7mm}p{15mm}p{11mm}p{13mm}p{15mm}p{41.3mm}p{46.5mm}ll}
\hline
\hline
Source           &Type  &Integral    &Variable                      &X-ray      &X-ray/optical/&No.of sources \\
                 &      &Pos. Unc.   &100s--1ks                     &Flux       &UV/IR/Radio   &in the X-     \\
                 &      &1.6$\sigma$ &$\frac{\rm IBIS}{\rm ISGRI}$  &15--40~keV &sources in X- &ray error circ.       \\
                 &      &            &                              &(mCrab)    &ray error circ.   &       \\
\hline
IGR J00370$+$6122  &HMXB$^{1,2}$   &2$'$        &Yes         &-         &BD$+$6073                     &\\
IGR J01363$+$6610  &HMXB$^{2}$       &2$'$        &Yes         &17        &HD9603                        &\\
IGR J16167$-$4957  &-      &6$'$$'$     &-           &2         &-                             &\\
IGR J16195$-$4945  &HMXB(?)$^{*,10}$   &16$'$$'$    &-           &-         &HD146628                      &\\
IGR J16207$-$5129  &-      &2$'$        &-           &-         &HD146803                      &\\
IGR J16358$-$4726  &LMXB(?)$^{3,4}$   &0.6$'$$'$   &Yes         &20-50     &2MASS J163553$-$472539        &\\
                   &Pulsar$^{5}$ &            &            &4.63      &                              &\\
IGR J16393$-$4643  &HMXB(?)&2$'$        &Yes         &3         &                           &20$^{**}$\\
                   &Pulsar$^{5,6,7,11}$ &            &            &          &                              &\\
IGR J16558$-$5203  &-      &8$'$$'$     &-           &-         &1RXS J165605$-$520345         &\\
                   &       &            &            &          &USNO-B1.0 0379$-$00008129     &\\      
IGR J17195$-$4100  &-      &8$'$$'$     &Yes         &-         &1RXS J171935$-$410054         &\\
                   &       &            &            &          &USNO-B1.0 0489$-$00511283     &\\
IGR J17200$-$3116  &-      &9$'$$'$     &Yes         &-         &1RXS J172006$-$311702         &\\
IGR J17252$-$3616  &HMXB$^{8}$&2$'$        &-           &-         &IRAS 17220$-$3615             &\\
                   &Pulsar &            &            &          &NVSS J172510$-$361614         &\\
                   &       &            &            &          &HD319824                      &\\
IGR J17254$-$3257  &-      &14$'$$'$        &-           &-         &1RXS J172525$-$325717         &\\
                   &       &            &            &          &USNO-B1.0 0570$-$00727635     &\\
IGR J17285$-$2922  &XB$^{9}$&2$'$       &Yes         &-         &IRAS 17252$-$2922             &\\
                   &       &            &            &          &[T66b]320                     &\\
IGR J17303$-$0601  &LMXB   &7$'$$'$        &Yes         &-      &H1726$-$058                   &\\
                   &       &            &            &          &USNO$-$A2.0 0825$-$10606993   &\\
                   &       &            &            &          &1RXS J173021.5$-$055933       &\\
IGR J17456$-$2901  &-      &1$'$        &-           &-         &1LC G359.923$-$00.013   &90$^{**}$ \\
IGR J17460$-$3047  &-      &2$'$        &-           &-         &                        &80 IR sources$^{**}$\\
IGR J17464$-$3213  &BHC$^{*}$&0.5$'$    &Yes         &60        &H1743$-$322                   &\\
                   &LMXB$^{3}$&         &            &          &2MASS 17461525-3213542 &\\
                   &          &         &            &          &USNO-A2.0 0525-294112269         &\\
IGR J17475$-$2822  &Sgr B2$^{5}$&2--3$'$    &-           &-         &                     &200 $^{**}$\\
IGR J17488$-$3253  &-      &12$'$$'$    &Yes         &-         &1RXS J174854.7$-$325444       &\\
IGR J18027$-$2016  &Pulsar$^{6}$ &1$'$        &-           &4.06      &HD312525                      &\\
                   &       &            &            &          &1LC G000.683$-$0.035          &\\
                   &       &            &            &          &IRAS 17594$-$2021             &\\
IGR J18406$-$0539  &-      &2--3$'$    &-           &-          &IRAS 18379$-$0546              &\\
                   &       &            &            &          &AX J1840.4$-$0537         &\\
                   &       &            &            &          &NVSS J184037$-$054317         &\\
                   &       &            &            &          &GSC2.2&&\\
IGR J18450$-$0435  &-      &2--3$'$    &-           &-          &IRAS 18422$-$0437              &\\
                   &       &            &            &          &PMN J1845$-$0433              &\\
IGR J18490$-$0000  &-      &2--3$'$    &-           &-         &-                              &\\

\hline
\end{tabular}
\footnotemark{High mass X-ray binary$^{1}$, Reig {et al.} 2005$^{2}$, Bird {et al.} 2004$^{3}$, Low mass X-ray binary$^{4}$, 
Revnivtsev {et al.} 2004$^{5}$, Lutovinov {et al.} 2005$^{6}$,Boudaghee {et al.} 2005$^{7}$, Zurita {et al.} 2005$^{8}$, 
 X-ray binary$^{9}$, Sidoli {et al.} 2005$^{10}$, Soldi {et al.} 2005$^{11}$  
http://isdc.unige.ch/~rodrigue/html/igrsources.html$^{*}$, $^{**}$ field sources identified at different wavelengths}
\end{center}
\end{table*}

\section{Observations and Analysis}
The radio observations were carried out at 0.61 GHz with bandwidths 
of 16 and 32 MHz respectively using the GMRT. The full array synthesized beam of the GMRT
antenna at 0.61 GHz is $\sim$ 5$'$. 

As described in paper-I (Pandey {et al.} 2005b), the association
of the possible radio counterpart to the {\emph{INTEGRAL}} source was based on the observed
radio morphology, flux density variability and cross-correlation with the NED catalogue.
During the observations in the Galactic center region at low radio frequencies, the gain decrease
due to system temperature becomes significant, hence system temperature corrections were applied
to the flux densities of the radio measurements.
The flux density scale was set by observing the primary calibrators 3C286, 3C147 and 3C48.
The phase calibrators were observed near the target source for $\sim$ 5 min scan interleaved with 25 min
scans on {\emph{INTEGRAL}} sources. The data recorded from GMRT was
converted into FITS files and analyzed using Astronomical Image Processing System (AIPS).
The details of observation and analysis procedure are desribed in detail in paper-I.
\par 
During our observations, radio sources were detected in the field of sixteen hard X-ray sources and a 
possible radio counterpart for the seventeenth source been detected during VLA observations. All the 
possible seventeen radio counterparts of the twenty three hard X-ray sources are reported in this paper 
and for remaining sources no
possible radio counterpart was detected within the 3$\sigma$ position error circle of the X-ray source.
In Table 1, we have listed physical properties of the sources observed during our survey along with the other
available data, inferred class of the object and the field sources in other wave bands within the
{\emph{INTEGRAL}} error circle.
A summary of the results of GMRT observations is given in Table 2. The best-fit radio positions for the
counterpart and the position offset with respect to the  X-ray position is given in column 7. Column 4
gives the radio flux density for the point (peak) and extended (total) sources. The rms
noise given in column 5 of the table corresponds to the average background  noise in the
image field  and  is higher in the Galactic plane. In the radio images presented below, the bold
ellipse marked with `A' indicates the radio counterpart close to the X-ray source.
We have grouped sources based on their radio morphology into point and extended sources.
We discuss in detail some of these sources with genuine radio -- X-ray association.

\section{Results}
\subsection{ Point radio sources within the field of {\emph{INTEGRAL}} sources:}
\noindent{\bf 1- IGR J17303$-$0601 :}
This source was detected in the Norma arm region (Walter {et al.} 2004) and in coincidence with the {\emph{ROSAT}} source 
1RXS J173021.5$-$055933 (Voges {et al.} 1999, Stephen {et al.} 2005). Two optical objects with $R \sim 15.5$ and $R \sim 18$ 
are found within the {\emph{ROSAT}} error box of 7$'$$'$ in the Digitized Sky Survey (DSS) field
( Monet {et al.} 2003, Stephen {et al}. 2005). Two near-infrared sources in the (2-MASS survey) 
are also coincident with the optical sources. The optical spectra of both these sources acquired with 
the Bologna Astronomical Observatory show features identical to
other X-ray emitting objects (Masetti {et al.} 2004).
The brighter source within the  {\emph{ROSAT}} error box was considered to be the optical counterpart to
IGR J17303$-$0601. All the optical emission lines of this object are at red shift zero, indicating its
Galactic origin. The presence of the He II line strongly indicates that this object
is undergoing mass accretion onto a compact star (e.g. van Paradijs \& McClintock 1995), 
thereby suggesting the  X-ray source to be  a low mass X-ray binary with IGR J17303$-$061 as the optical 
counterpart. The optical photometry of the source however, suggests the hard X-ray source to be an
intermediate polar with a spin period of 128 s (Gansicke et al. 2005).

\begin{figure}[h]
\begin{center}
\psfig{figure=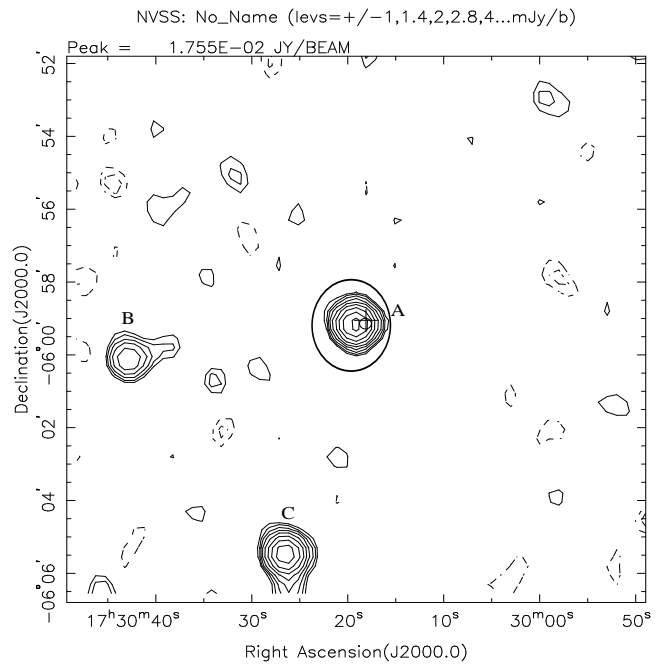,width=8.5cm,height=8.7cm,angle=0}
\end{center}
\caption{\footnotesize{NVSS image of IGR J17303$-$0601 at 1.4 GHz with the
{\emph{INTEGRAL}} position marked with  $+$. Source `A' is coincident in position with IGR J17303$-$0601. 
Sources `B' and `C' are field sources. The circle shows the {\emph{INTEGRAL}} uncertainty error 
circle of 3$\sigma$ (14$'$$'$)}}
\label{fig.1.}
\end{figure}

Radio observations of the source were made using GMRT, with the aim of finding any possible radio counterpart within the location error box of the X-ray source however no flux was detected  from the source as seen in  Table I.   We then analysed the NVSS observations or the region conducted at 1.4 GHz with the VLA (Condon {et al.} 1998) and the  radio image of the field is shown in Fig. 1. 
A compact point source `A' of $\sim$18 mJy at 1.4 GHz is detected coincident with
the {\emph{INTEGRAL}} and {\emph{ROSAT}} source position. The precise position with the radio observations
is RA: 17h 30m 21.50s and DEC: $-$05d 59$'$ 33.5$'$$'$ (J2000). Two other compact field sources `B' with radio
flux density $\sim$5 mJy and RA: 17h 30m 45.48 and DEC: $-$06d 01$'$ 57.17$'$$'$ and `C' with radio flux
density $\sim$16 mJy and RA: 17h 30m 28.41 and DEC: $-$06d 05$'$ 56.27$'$$'$, were clearly detected within
the field and lie outside the {\emph{INTEGRAL}} error circle.
In order to search for the radio source at lower frequencies GMRT observations on this source
was performed at 0.61 GHz. No radio source was found coincident in position with the NVSS source
`A' during our observations at 0.61 GHz. Fig.1 shows the NVSS field image of the source IGR J17303$-$0601.
The 3$\sigma$ upper limit in the GMRT image was $\sim$6 mJy and the rms
noise was $\sim$1.84 mJy beam$^{-1}$. The sources `B' and `C'  with radio flux density $\sim$18 mJy
and $\sim$15.5 mJy were clearly detected coincident in position with the NVSS sources.

At radio wavelengths the detection at 1.4 GHz and non-detection at
0.61 GHz with positive detection of other field sources suggest that IGR J17303$-$0601 is a
radio emitting X-ray binary (REXB). And the source is highly variable or absorbed at low frequencies. The
absorption may be due to synchrotron self absorption process dominant at low frequencies as seen for most
of the  X-ray binaries (Pandey {et al.} 2004). In both the cases, a non-thermal origin of the radio emission is
favored.
Like few other sources in the Norma Arm region
even this source was highly absorbed at low frequencies (Pandey {et al.} 2005b).\\\\
\noindent{\bf 2- IGR J17464$-$3213 :}
This transient BH candidate was detected by the {\emph{INTEGRAL}} satellite on 21st March, 2003 
(Revnivtsev {et al.} 2003). The position of the source is consistent with the position of the HEAO source H1743$-$322 
(Markwardt \& Swank 2003, Gursky {et al.} 1978). The spectral fit to the JEM-X/IBIS data shows the presence of 
a soft component fitted by a multi color disk black body 
and a hard power-law tail with photon index of 2.2 which extends to 80 keV. The observations made 
during the outburst shows that the light curve is typical of a X-ray nova 
(Steeghs {et al.} 2003). 
It is therefore believed that IGR J17464$-$3213 is a classical X-ray nova -- a
LMXB harboring a BH -- which experienced a recurrent outburst in 2003 (Lutovinov {et al.} 2005).

The RXTE pointed observations on 28th Mar, 2003 gave mean fluxes 50, 
200 and 220 mCrab in 2 -- 10, 15 -- 40 and 40 -- 100~keV range respectively. A strong quasi periodic 
oscillation (QPO) with the period 
$\sim$ 20 s was also seen in the X-ray light curve. The X-ray spectrum is consistent with an 
absorbed power law with photon index 1.49$\pm$0.01 and an absorption column of 2.4 $\times$ 
10$^{22}$ cm$^{-2}$. Compton reflection signatures 
are seen in the continuum spectrum (Grebenev {et al.} 2003, Capitanio {et al.} 2005).  
\begin{figure}[t]
\begin{center}
\psfig{figure=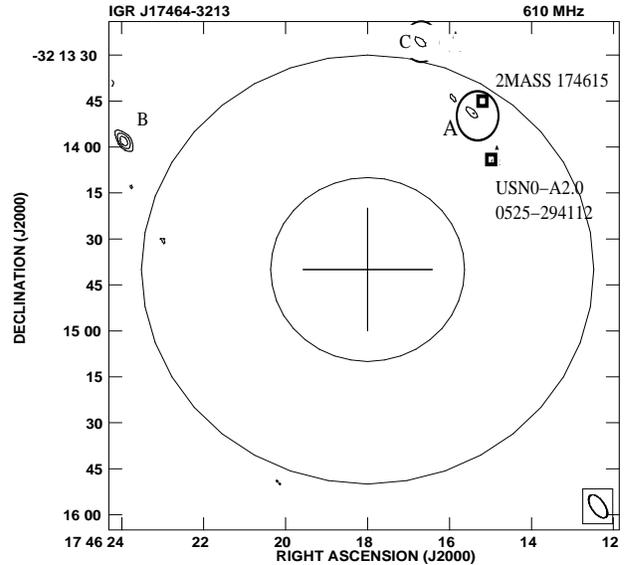,width=8.5cm,height=8.7cm,angle=0}
\end{center}
\caption{\footnotesize{GMRT image of IGR J17464$-$3213 at 0.61 GHz with the
{\emph{INTEGRAL}} position marked with  $+$
. The small and large circles shows the
{\emph{INTEGRAL}} uncertainty error circles of 1.6$\sigma$ and 3$\sigma$. The contour levels are 
2 mJy $\times$ 1, 1.4, 2}. The field sources 2MASS 17461525-3213542 and USNO-A2.0 0525-294112269 
are marked in the figure.}
\label{fig.1.}
\end{figure}
The RXTE monitoring of the source between May -- July,  2004 observed IGR J17464$-$3213 in several 
BH states and revealed various types of variability, including QPOs of 7.8 Hz (Homan {et al.} 2005).

During the follow up observations with the VLA on 30th, March and 1st April, 2003 
a compact, variable source was detected at 4.8 GHz at, RA: 17h 46m 15.61 $\pm$ 0.01s, 
DEC: $-$32d 13$'$ 59.9 $\pm$ 1.0$'$$'$ (J2000) and 
approximately 0.64$'$ from the original {\emph{INTEGRAL}} position. 
It is consistent with the position of H1743$-$322 (Swank {et al.} 2004). The source flux was 4 mJy 
on 30th March, and had brightened by about 50\% on 1 April. A strong radio flare 
was detected on 8th April, 2003 (Rupen {et al.} 2003a). The ATCA radio observations of H1743$-$322 
performed from Nov, 2003 -- June, 2004 led to the discovery of large-scale radio jets on each side 
of the BHC H1743$-$322 (Corbel {et al.} 2005).

The optical observations in the $I$-filter at the radio position show a marginal detection of an 
optical counterpart at a level of $I\sb{mag} \sim 20$ (Khamitov {et al.} 2003, Remillard {et al.} 
2003, Steeghs {et al.} 2003, Rupen {et al.} 2003b). 

On 3rd Jul, 2004 a second outburst was detected in the X-ray light curve of IGR J17464$-$3213 and it went
into hard X-ray state (Swank {et al.} 2004). During the end of the X-ray state, the radio
source was not detected at the position mentioned by (Rupen {et al.} 2003a) until 4th Aug 2004 during
VLA observations at 4.86 GHz; however,
it was positively detected by VLA on 5th Aug 2004, with a flux density of 1.96$\pm$0.15 mJy
(Rupen {et al.} 2004). During GMRT observations on 25th July, 2004, a radio source was detected coincident with 
the VLA position reported by Rupen {et al.} 2003a, at a radio flux density of 2.75$\pm$0.52 mJy refer Fig. 2. 
It is also interesting to note that two compact sources, (B) of flux level 6 mJy at position coordinates, RA: 17h 46m 24.17s
and DEC: $-$32d 13$'$ 59.92$'$$'$ and (C) of flux level 2 mJy at position coordinates, RA: 17h 46m 16.55s
and DEC: $-$32d 13$'$ 29.89$'$$'$ are also detected within the field of IGR J17464$-$3213; however, they clearly 
lie outside the {\emph{INTEGRAL}} position error circle. The analysis of the NVSS data at 1.4 GHz shows no 
point sources coincident with the 
radio sources (A) and (B), detected by the GMRT; however, the source (C) was positively detected in the NVSS field at a 
flux level of $\sim$1.05 mJy. There were no other radio observations reported at this epoch. 
Hence two variable radio sources (A, B) are detected witin the field of {\emph{INTEGRAL}} source; however, the source (A) is 
most likely to be associated with the hard X-ray source and it clearly shows transient behavior at radio wavelengths 
like other LMXBs. 

In order to determine IR and optical counterparts, we have analysed the 2MASS and USNO data near the radio source (A) position 
for the field of IGR J17464$-$3213. An IR point source, 2MASS 17461525-3213542 with coordinates RA: 17h 46m 15.25s
and DEC: $-$32d 13$'$ 54.2$'$$'$ and magnitudes, $J= 16.2, H= 13.8, K= 13.16$ and an optical source, 
USNO$-$A2.0 0525$-$294112269 with position coordinates RA: 17h 46m 14.77s
and DEC: $-$32d 14$'$ 06.02$'$$'$ and magnitudes, $B= 19.6, R= 17.4$ lies within the error circle of the hard 
X-ray source and close to the radio source (Voges {et al.} 2004). Thus H1743$-$322, 2MASS 17461525-3213542, and 
USNO$-$A2.0 0525$-$294112269, may be most likely associated with IGR J17464$-$3213.

Thus possible counterparts in IR and optical wavelengths and a variable radio counterpart for the source 
IGR J17464$-$3213 has been detected, which confirms that IGR J17464$-$3213 is a REXB 
and a microquasar candidate. However, it is necessary to 
perform simulataneous radio observations during the flaring episodes to understand the system in detail.   
\\\\
\noindent{\bf 3- IGR J18406$-$0539 :}
The source was discovered during the observations of the Sagittarius arm region by IBIS telescope during the spring
of 2003 (Belanger {et al.} 2004). We have carried out cross identification of this source with the data available 
from various catalogues, to identify the nature of the source. A hard X-ray source, 
AX J1840.4$-$0537 discovered during ASCA observations with RA: 18h 40m 24s 
and DEC: $-$05d 37$'$ 00$'$$'$ lies within the position error circle of the source  
(Bamba {et al.} 2003). The field is further complicated by the 
presence of an optical source, GSC2.2 with magnitudes $B=16$ and $R=19$ and position 
RA: 18h 40m 38.094s and DEC: $-$05d 43$'$ 19.30$'$$'$ lying within error circle (Monet {et al.} 1998). 
Fig. 3 shows the radio image of the field of IGR J18406$-$0539 at 0.61 GHz with the other known field sources .

An IR point source, 1RAS J18379$-$0546 (Cutri {et al.} 2003), with coordinates RA: 18h 40m 38.04s, DEC: $-$05d 43$'$ 20$'$$'$
and magnitudes $J=12.89$, $H=11.91$, $K=11.61$, also lies within the 2$\sigma$ position error ellipse of the hard X-ray 
source. 
Thus the sources AX J1840.4$-$0537, GSC2.2 and 1RAS J18379$-$0546 are
therefore, likely associated with the counterparts of IGR J18406$-$0539.
\begin{figure}[t]
\begin{center}
\psfig{figure=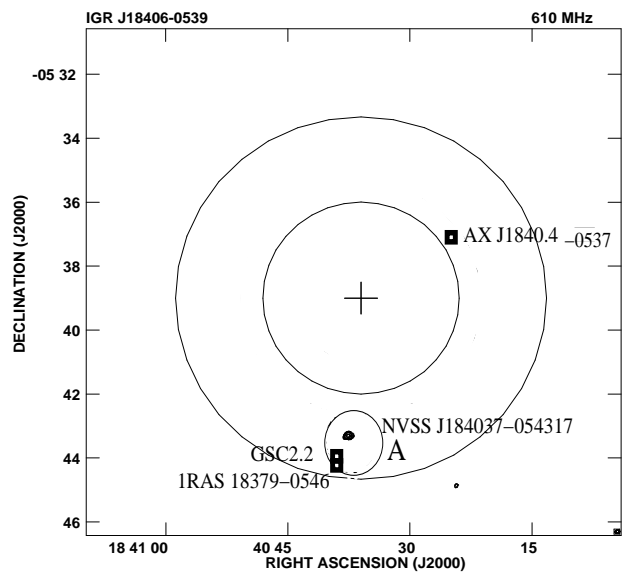,height=8.5cm,width=8.5cm,angle=0}
\end{center}
\caption{\footnotesize{GMRT image of IGR J18406$-$0539 at 0.61 GHz with the
{\emph{INTEGRAL}} position marked with  $+$
. The small and large circles shows the
{\emph{INTEGRAL}} uncertainty error circles of 1.6$\sigma$ and 3$\sigma$. The contour levels are 
5.5 mJy $\times$ 1, 2, 4, 8. The boxes show the ASCA, optical and IR sources within the {\emph{INTEGRAL}} error circle. 
The field sources AX J1840.4$-$0537, 1RAS 18379$-$0546, NVSS J184037$-$054317 and GSC2.2 are marked in the figure.}}
\label{fig.1.}
\end{figure}
\begin{figure}[t]
\begin{center}
\psfig{figure=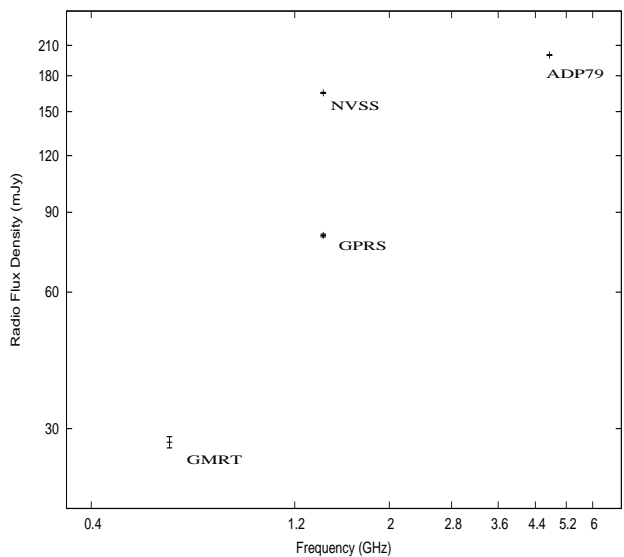,height=7.5cm,width=8.5cm,angle=0}
\end{center}
\caption{\footnotesize{Radio spectrum for IGR J18406$-$0539}}
\label{fig.1.}
\end{figure}

During GMRT observations a point source was detected with radio flux density of $\sim$ 28 mJy and at RA: 18h 40m 37.61s and
DEC: $-$05d 43$'$ 17.99$'$$'$, which is 4.31$'$ away from the hard X-ray source. The
NVSS image of the field also shows a point source of 165 mJy coincident with the GMRT position.
Using the various radio survey, we have computed the radio spectrum for the source (non-simultaneous measurement) in Fig. 4.
It can be seen from the figure that the spectrum is highly inverted at meter wavelengths. 
It is interesting to note that the NVSS survey at 1.4 GHz and the Galactic Plane Radio-source 
Survey at 1.4 GHz (Zoonematkermani {et al.} 1990) for the compact Galactic source at different epochs
gave the radio flux density of 165 mJy and 80 mJy respectively. A variability in the
radio flux density by a factor of two is clearly measured. The Galactic plane survey
at 4.87 GHz by (Altenhoff {et al.} 1979) gave the radio flux density for the source
as 200 mJy. This information implies that the source shows absorption at
lower frequencies and is variable in nature. It is important to do further
observations in the optical and infrared band to confirm the nature of the
companion star. However, from the available information we infer that the source
IGR J18406$-$0539 is most likely a REXB, transient in nature and a possible microquasar candidate.
\subsection{Extended radio sources within the field of {\emph{INTEGRAL}} sources:}
\subsubsection{Extragalactic radio sources within the field of {\emph{INTEGRAL}} sources:}
The Galactic binary X-ray sources are mostly unresolved at arcsec scale (Pandey {et al.} 2005a). The
extended sources can thus be catagorized as extragalactic in nature. From our GMRT observations, we find
that {\bf among the remaining twenty sources,} the radio emission observed in the direction of three sources, viz,
IGR J17195$-$4100, IGR J17200$-$3116 and IGR J17456$-$2901 is extended in nature and having a double 
source morphology, which is typical of the extragalactic sources.
\begin{figure}[b]
\begin{center}
\psfig{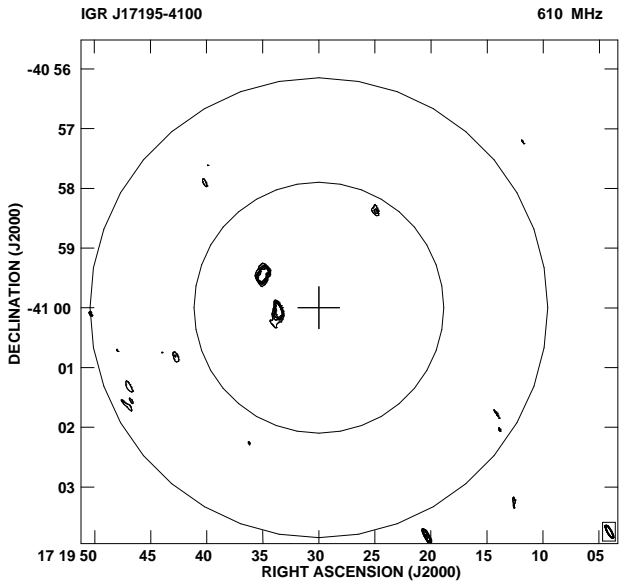}
\end{center}
\caption{\footnotesize{GMRT image of IGR J17195$-$4100 at 0.61 GHz with the
{\emph{INTEGRAL}} position marked with  $+$
. The small and large circles shows the
{\emph{INTEGRAL}} uncertainty error circles of 1.6$\sigma$ and 3$\sigma$. The countour levels are
2.3 mJy $\times$ 1.0, 1.2, 1.3, 1.4, 2.0, 2.8, 4.0, 8.0, 16.0, 32.0}}
\label{fig.1.}
\end{figure} 
Hence, we classify these as extragalactic radio sources. In Fig. 5, we have plotted the GMRT image of 
IGR17195-4100 taken at 0.61 MHz. The data clearly suggests the sources to be a  a radio galaxy.
The cross identification with the NED catalogue (Voges {et al.} 1999, Condon {et al.}
1982, Pappa {et al.} 2001) also confirms their extragalactic nature of both these regions.
In regards to the true association of the radio source with the hard X-ray source, it is quite likely that the 
X-ray sources are indeed extragalactic. No X-ray spectral data is yet available to infer the galactic origin of 
the source. However, if the X-ray sources are galactic in origin, then the observed radio association will be a case of 
fortutious line of sight coincindence.  we have listed the known field sources within the {\emph{INTEGRAL}} error circle 
in Table 1. The optical counter part for IGR J17195$-$4100  in the table is listed from USNO catalogue.

\subsubsection{ Extended Galactic radio sources in diffuse region within the field of {\emph{INTEGRAL}} sources:} 
{\bf Among the rest seventeen sources, eleven hard X-ray sources do show extended radio morphology within the  position error 
circle. The available X-ray spectra on of few of these sources suggest their binary nature and galactic origin 
(refer table 1). No known extragalactic radio sources from the NED catalogue lie in the position
error circle of these {\emph{INTEGRAL}} sources and in coincidence with GMRT position. The observed radio emission with morphology
non similar to the radio galaxies and with no known extragalactic identification therefore suggests that these extended 
sources may most probably be associated with the radio emitting regions within the galaxy.}
\begin{figure}[b]                                                                                                                    \begin{center}                                                                                                                       \psfig{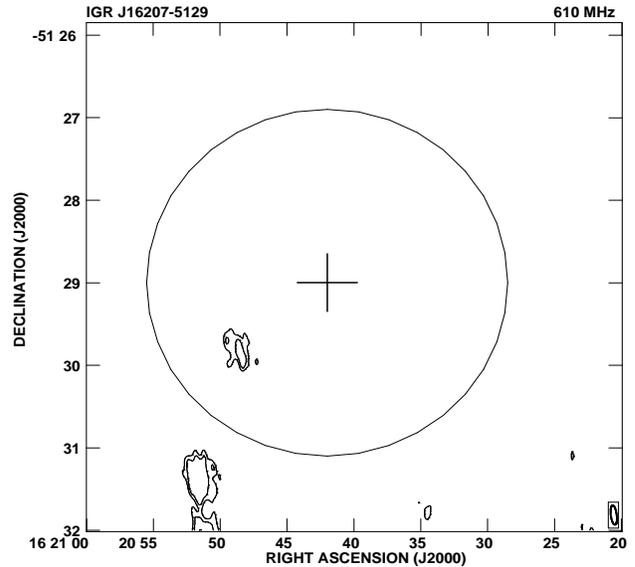}
\end{center}
\caption{\footnotesize{GMRT image of IGR J16207$-$5129 at 0.61 GHz with the
{\emph{INTEGRAL}} position marked with  $+$
. The circle shows the {\emph{INTEGRAL}} uncertainty error circle of 1.6$\sigma$. The countour levels are
3.0 mJy $\times$ 1.0, 1.2, 2.0, 2.8, 4.0, 8.0, 16.0, 32.0.}}
\label{fig.1.}
\end{figure}
From the GMRT data, based on their radio morphology, the eleven {\emph{INTEGRAL}} hard X-ray sources 
can be grouped in two classes; (a) extended Galactic source and (b) extended source typical of the diffuse emission regions.
\begin{figure}[h]
\begin{center}
\psfig{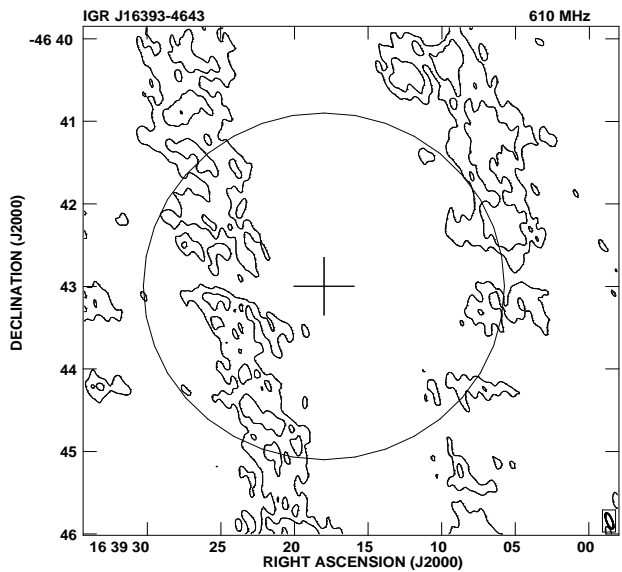}
\end{center}
\caption{\footnotesize{GMRT image of IGR J16393$-$4643 at 0.61 GHz with the
{\emph{INTEGRAL}} position marked with  $+$
. The circle shows the {\emph{INTEGRAL}} uncertainty error circle of 1.6$\sigma$. The countour levels are
2.0 mJy $\times$ 1, 2, 4, 18, 16, 32, 64}}
\label{fig.1.}
\end{figure}
In the Fig. 6 and 7 we have plotted representative radio images for the two groups. The image shown in 
figure 6 is similar to the one having an extended jet emission  while figure 7 is reminscent of the molecular clouds.
\par
Five hard X-ray sources, namely, IGR J16207$-$5129, IGR J16558$-$5203, IGR J17285$-$2922, IGR J17460$-$3047 and 
IGR J18450$-$0435 belongs to group (a). Except IGR J17285-2922, all sources are located in the Norma Arm region 
and have high probability of being galactic in nature. The X-ray spectrum of IGR J17285$-$2922 shows XB 
characteristics (Barlow {et al.} 2004); however, the nature of remaining sources is not yet identified.
Positive detection of extended radio emission at low frequencies, associated with the galactic X-ray sources seen in
the GMRT data suggests the presence of a new class of galactic extended radio sources.
\par
Six sources, viz IGR J16167$-$4957, IGR J16195$-$4945,
IGR J16393$-$4643, IGR J17252$-$3616, IGR J17254$-$3257, and
IGR J17475$-$2822 belongs to group (b). Except IGR J17285$-$2922, all the sources in this group were detected 
in the Norma Arm region {\bf (refer table 1)}. 

As seen from  Fig. 7, due to the large extent of the radio 
emission, it is difficult to associate a single region with the X-ray source, even though the radio 
emission lies within the position error circle of the X-ray source. 
The sources IGR J16393$-$4643, 
IGR J17456$-$2901, IGR J17460$-$3047 and IGR J17475$-$2822 have crowded fields with large number of 
field sources $>$20,  as expected from the diffuse regions. Therefore, we conclude  none of these radio sources
may be associated with the {\emph{INTEGRAL}} sources.

\subsection{X-ray sources with no radio counterpart}
Six sources viz, IGR J00370$+$6122, IGR J01363$+$6610, IGR J16358$-$4726, IGR J17488$-$3253,
IGR J18027$-$2016 and IGR J18490$-$0000 have no GMRT counterpart. It has been suggested that 
IGR J00370$+$6122 and IGR J01363$+$6610 are HMXB systems (Reig {et al.} 2005). Hence either these sources 
are not REXBs or these are synchrotron self absorbed at low frequencies at which our observations were made or 
these are highly variable in radio band. Follow up observations in the radio window is necessary to 
confirm the variable  nature of these sources.

\section{Summary and Conclusion}
We have presented the radio analysis of 23 of the newly discovered {\emph{INTEGRAL}} 
hard X-ray sources. Most of these sources are X-ray binaries; however, the 
identification of AGNs, radio galaxies, X-ray novas, CVs and pulsars are the other 
important byproducts. Among the twenty three sources observed, seventeen have a 
possible radio counterpart detected at radio wavelengths. The position offset 
of the possible radio counterparts with respect to the {\emph{INTEGRAL}} position is 
of the order of few arc minutes. The consistent position provided by the GMRT will 
allow the search for infrared/optical counterparts for these sources to be detected. 
Based on the radio morphology of these source we have further grouped them into:\\
(a) Galactic point source,\\ 
(b) extended Galactic sources and sources in diffuse Galactic emission,\\
(c) extragalactic sources.\\ 
Three sources belong to group (a) and are REXBs. We carried out a detailed study 
of these three sources and their possible counterparts at other wavebands. Based on the 
variability in the radio and X-ray windows along with the information available about 
the counterparts in other wavebands we infer that IGR J17303$-$0601, IGR J17464$-$3213 and 
IGR J18406$-$0539 are possible microquasar candidates. 
However, devoted radio observations 
are necessary to confirm the jet emission from these sources. We have also detected a variable compact radio
source within the field of IGR J17464$-$3213. Eleven sources can be associated 
with group (b), the diffused Galactic region, and three sources satisfy the radio morphology 
of extragalactic sources, group (c). No radio counterpart was detected for the remaining sources. 

To conclude, we highlight that our observations were very important in pinpointing the 
possible microquasar candidates from the list of 40 {\emph{INTEGRAL}} sources observed 
at radio wavelength. We have performed repeated observations on these sources of our 
interest in Cycle 7 to look for the radio variability and the data has to be analyzed.

\begin{acknowledgement}
I would also like to thank Dr. M. Ribo, Dr. D. Green and Dr. D. Ojha for the useful discussions and Prof. V. Kulkarni, 
Dept. of Physics, Mumbai University for his constant support and suggestions.

\end{acknowledgement}

\cite{whatever}
\begin{table*}
\begin{center}
\caption{Possible Radio counterparts of target INTEGRAL sources observed with GMRT at 0.61 GHz}
\begin{tabular}{llllllllll}
\hline
\hline
Source    &Date    &S$\sb{\nu}$   &$\sigma$       &Radio     &Pos. Off.&Radio    &S$\sb{\nu}$(NVSS)\\
          &dd/mm/yy&(Peak/Total)  &(mJy b$^{-1}$) &Pos. GMRT &w.r.t    &Structure&(Peak/Total)\\
          &        &(mJy)   &               &RA \& DEC &Integral Pos.&GMRT &(mJy)\\
\hline
IGR J00370$+$6122 &25/06/04&$\le$7.00     &2.18  & -                        &-      &-          &$\le$1.6\\
IGR J01363$+$6610 &25/06/04&$\le$7.00     &2.31  & -                        &-      &-          &2.5(E)\\
IGR J16167$-$4957 &23/07/04&725           &2.90  &16h 16m 44.29s$\pm$0.91   &0.12$'$&Extended   &N/A\\
                  &        &              &      &-49d 57$'$ 10.02$'$$'$$\pm$0.71   && (DG)     &    \\
IGR J16195$-$4945 &30/07/04&256.4         &0.84  &16h 19m 35.07s$\pm$0.21   &0.84$'$&Extended   &N/A\\
                   &       &              &      &-49d 44$'$ 59.01$'$$'$$\pm$0.28   && (DG)     &    \\
IGR J16207$-$5129 &30/07/04&60.50         &0.78  &16h 20m 48.54s$\pm$0.77   &1.33$'$&Extended   &N/A\\
                    &      &              &      &-51d 29$'$ 50.01$'$$'$$\pm$0.98&  &(DG)       &    \\
IGR J16358$-$4726 &31/07/04&$\le$7.50     &2.5   &-                         &-      &-          &N/A\\
IGR J16393$-$4643 &23/07/04&79.25         &1.63  &16h 39m 03.9s$\pm$0.51    &2.58$'$&Extended   &N/A\\
                    &      &              &      &-46d 42$'$ 15.55$'$$'$$\pm$0.59&  &(DG)       &    \\
IGR J16558$-$5203 &30/07/04&27.24         &0.78  &16h 55m 46s$\pm$0.04      &0.69$'$&Extended   &N/A\\
                    &      &              &      &-52d 03$'$ 58$'$$'$$\pm$1.04&     & (E)       &     \\
IGR J17195$-$4100 &23/07/04&33.44         &0.56  &17h 19m 34s$\pm$0.22      &0.98$'$&Extended   &N/A\\
                    &      &              &      &-41d 00$'$ 00$'$$'$$\pm$1.24&     &(DS-E)     & \\
IGR J17200$-$3116&23/07/04 &33.03          &0.51  &17h 19m 55s$\pm$0.74     &0.55$'$&Extended   &20 (E)\\
                    &      &              &      &-31d 16$'$ 01$'$$'$$\pm$0.70&     &(DS-E)     & \\
IGR J17252$-$3616&25/07/04 &36.74         &1.88  &17h 25m 11.09s$\pm$1.71   &0.82$'$&Extended   &700 (E)\\
                    &      &              &      &-36d 16$'$ 48.01$'$$'$$\pm$0.20&  &(DG)       &    \\
IGR J17254$-$3257&25/07/04 &359.65        &0.43  &17h 25m 24s$\pm$1.22      &3.12$'$&Extended   &$\le$2.5\\
                    &      &              &      &-32d 55$'$ 10$'$$'$$\pm$1.33&     &(DG)       &    \\
IGR J17285$-$2922&25/07/04 &74.88         &0.59  &17h 28m 28.75s$\pm$0.07   &0.96$'$&Extended   &$\le$2.7\\
                   &       &              &      &-29d 21$'$ 04.51$'$$'$$\pm$1.24&  & (E)       &    \\
IGR J17303$-$0601&25/06/04 &$\le$6.00     &1.84  &-                         &-      &-          &18 (P)\\
IGR J17456$-$2901&23/07/04 &23.11         &1.53  &17h 45m 38.07s$\pm$0.04   &0.56$'$&Extended   &17200 (E)\\
                    &      &              &      &-29d 00$'$ 40.00$'$$'$$\pm$1.87  &&(DS-E)     &  \\
IGR J17460$-$3047&30/07/04 &2.50          &0.46  &17h 45m 59.69s$\pm$0.03   &0.076$'$&Extended  &$\le$2.3\\
                    &      &              &      &-30d 46$'$ 58.00$'$$'$$\pm$2.33  & & (E)      &      \\
IGR J17464$-$3213&25/07/04 &2.75$\pm$0.52 &0.50  &17h 46m 15.61s$\pm$0.28   &0.64$'$&Point      &$\le$2.4\\
                 &         &              &      &-32d 13$'$ 59.9$'$$'$$\pm$0.21    &           &   \\
IGR J17475$-$2822&23/07/04 &12.50         &1.27  &17h 47m 25.68s$\pm$0.62   &1.02$'$&Extended   &1920 (E)\\
                    &      &              &      &-28d 22$'$ 21.96$'$$'$$\pm$1.21  &&(DG)       &        \\
IGR J17488$-$3253&30/07/04 &$\le$2.10     &0.67  &-                         &-      &-          &$\le$1.6\\
IGR J18027$-$2016&25/06/04 &$\le$7.00     &2.32  &-                         &-      &-          &$\le$1.3\\
IGR J18406$-$0539&23/07/04 &28.03$\pm$0.76&0.76  &18h 40m 37.61s$\pm$0.12   &4.31$'$&Point      &165 (P)\\
                    &      &              &      &-05d 43$'$ 17.99$'$$'$$\pm$0.18  &&           &     \\
IGR J18450$-$0435&23/07/04 &207.94        &0.76  &18h 45m 12.05s$\pm$1.06   &5.90$'$&Extended   &74 (E)\\
                    &      &              &      &-04d 40$'$ 05.99$'$$'$$\pm$1.69  && (E)       &$\le$1.4       \\
IGR J18490$-$0000&23/07/04 &$\le$3.5      &1.25  &-             &-      &-                &     &\\
\hline
\end{tabular}
\footnotemark{~~~~~DS-E: Double Source Extragalactic, E: Extended, P: Point, DG: Diffused Galactic Region}
\end{center}
\end{table*}

\end{document}